\documentclass[prb,aps,floatfix,showpacs,twocolumn]{revtex4}
\usepackage[intlimits]{amsmath}
\usepackage{amsfonts}
\usepackage[T1]{fontenc}
\usepackage{array}
\usepackage{graphicx}
\newcolumntype{C}{>{$}c<{$}}
\usepackage{dsfont}

\newcommand{\I}{\mathrm{i}}
\newcommand{\e}{\mathrm{e}}
\newcommand{\D}{\mathrm{d}}
\newcommand{\Q}{\mathbf Q}
\newcommand{\R}{\mathds R}
\newcommand{\id}{\mathds 1}

\newcommand{\detg}{{\operatorname{detg}}}
\newcommand{\trg}{{\operatorname{trg}}}
\newcommand{\tr}{{\operatorname{tr}}}

\newcommand{\erf}{{\operatorname{erf}}}

\renewcommand{\theta}{\vartheta}
\renewcommand{\epsilon}{\varepsilon}
\renewcommand{\bar}{\overline}
\renewcommand{\Re}{\operatorname{Re}}
\renewcommand{\Im}{\operatorname{Im}}

\begin{document}

\title{Exact Density of States for finite Gaussian Random Matrix
  Ensembles via Supersymmetry}

\author{ Frieder Kalisch\footnote{e-mail:
    frieder.kalisch@physik.uni-augsburg.de} and Daniel Braak}

\affiliation{Institut f\"ur theoretische Physik II, Universit\"at Augsburg,\\
  86135 Augsburg, Germany}

\date{January 31, 2002}

\begin{abstract}
We calculate the exact density of states (DOS) for the three classical and
two
non-classical Random Matrix Ensembles for finite matrix size $N$ using
supersymmetric integrals. The $1/N-$Expansion yields 
already in lowest order  good
approximations to the exact result even for small values of $N\sim 5$.
We conjecture a connection between the $N-$dependence of the oscillating
part of the DOS and
the short-distance behavior of the two-level correlation function.  
\end{abstract}

\pacs{02.50.Ey,05.45.+b, 21.10.-k, 24.60.Lz, 72.80.Ng}

\maketitle

\section{Introduction}
\label{sec:intro}
Random Matrix Theory (RMT) has become a well-established tool to 
describe the statistical properties of the energy levels of 
quantum many-particle systems in the ergodic regime
\cite{guhr}. It is based on the assumption that (apart from global
symmetry conditions as spin rotation- or time reversal invariance) 
the Hamiltonian $H$ of a system with $N$ states
is described by a probability distribution
\begin{equation*}
\langle H_{ij}\rangle=0, \qquad \langle
H_{ij}H^*_{kl}\rangle=g\delta_{ik}\delta_{jl}.
\end{equation*}
RMT derives the statistical properties of the eigenvalues of $H$
from this distribution of its matrix elements.
Whereas most of the applications are still centered
around the three classical Wigner-Dyson (WD) Ensembles, the Gaussian
unitary, orthogonal  and symplectic ensembles (GUE, GOE and GSE), there
is also interest in certain generalizations of the standard cases,
leading to seven additional symmetry classes, which differ from each
other
by their behavior under a set of discrete space-time transformations.
Three of them are the chiral analogues to the WD-Ensembles
and four are relevant for mesoscopic normal-superconducting
hybrid systems \cite{altz}. 

RMT is a soluble description of the statistics of disordered and
chaotic
systems in the sense that all $n-$level correlation functions can
be computed in principle exactly. The most important of these correlation
functions is the two-level correlation function as it contains
already the main information about the repulsion of neighboring
levels and discerns the three WD-Ensembles
unambiguously. Several mathematical tools are available to compute these
correlation functions, e.g.\ the method of orthogonal polynomials
\cite{mehta}, or the sypersymmetric method \cite{efetov}. The last
method maps the matrix ensemble to a zero-dimensional 
non-linear sigma-model
in 
the limit of large matrix size $N$. Because then the limit
$N\rightarrow\infty$ is implicit in all calculations, it is not
possible to study directly finite-size effects, i.e.\ the dependence
of the average level spacing $\Delta(E)$
at energy $E$ on $N$.
 
Instead one is forced to fix a certain scaling of the overall
normalization of $H$ with $N$. In this way the two energy scales, the
bandwidth $2E_{max}$ and the mean level spacing $\Delta(E)$, are
entangled with the matrix size.  Either the bandwidth is kept at a
fixed finite value (macroscopic scaling) or the mean level spacing
$\Delta(E) \sim 1$ at a certain energy $E$ (microscopic scaling). The
first scaling is used for the DOS (and yields the semi-circle for
$N\rightarrow\infty$), the second is appropriate to extract the
universal level repulsion.  The non-classical Ensembles (we will treat
class D and Class C in this article) show characteristic features of
the DOS, deviating from the classical ensembles, but this was
calculated only in the microscopic scaling
limit.\cite{altz,nagao,ivanov} The DOS varies strongly on the scale of
the level spacing for $E\sim 0$, and becomes therefore dependent on
the position within the spectrum.  In this case the usual unfolding
procedure which normalizes the DOS to a constant over many consecutive
levels is inapplicable. The question arises, whether the new features
of the DOS survive the macroscopic scaling $\Delta(E)\sim 1/N$ leading
to a finite band width, which is the physical relevant case for solid
state applications.  It seems therefore advisable to compute the DOS
exactly for finite $N$ and to study its behavior in the limit
$N\rightarrow\infty$ using this formula. The only results for finite
$N$ were so far obtained via the orthogonal polynomial method. We will
show in the following that the supersymmetric method can be used as
well while being technically simpler and more versatile. 

The outline of the paper is as follows: In section II we study the
simplest case, the GUE, and explain in detail how the exact DOS for
finite $N$ can be obtained.  We then use a saddle-point method to
derive a $1/N$-expansion. This expansion is not based on a
supersymmetric saddle-point manifold but to the contrary is effected
by exact integration over the fermionic variables, which inevitably
breaks explicit supersymmetry.  Taking into account the quadratic
fluctuations of the remaining bosonic variables yields then already
excellent approximations to the exact result for small $N$ and
everywhere in the spectrum except at the band edge, where the
$1/N-$expansion diverges.  In Section \ref{sec:three} we give the
exact DOS for finite $N$ of the orthogonal and symplectic ensembles as
well as class C and D.  These ensembles are especially simple among
the non-classical ones because the joint eigenvalue distribution can
be associated with free fermions on a line in an external
potential\cite{altz}. Nevertheless the calculations of the DOS become
only simple if the confining part of the potential is neglected,
corresponding to microscopic scaling, thereby loosing all information
about the relation between $\Delta(E)$ and $N$ which is just given by
the confinement. Therefore we treat Class C/D here together with the
WD-Ensembles. Regarding the level repulsion they belong to the unitary
universality class, i.e.\ without time reversal invariance (see below).  Our
method, of course, extends to all other Gaussian Ensembles as well.
Section \ref{sec:four} contains the $1/N-$expansion around the bosonic
saddle-points for GOE, GSE and Class C/D.  We observe a relation
between the $N-$dependence of the oscillatory part of the correction
terms and the short distance behavior of the two-level correlation
function. Then we determine the average level spacing in the band
center, where the DOS of class C and D deviates strongly from the
(constant) behavior known from the WD-Ensembles, in a way independent
from any scaling assumption.  Section \ref{sec:concl} contains a
summary of our results.

\section{The unitary Ensemble}
\label{sec:two}
The ensemble consists of complex hermitian $N\times N$ Matrices $H=H^\dagger$.
We represent the DOS at complex energy $z$ ($\Im z>0$) as the
following expectation value (summation convention is understood):
\begin{equation}
  \label{eq:1}
  \langle n(z) \rangle = \frac1{\pi N} \Re \int\!\! \mathcal{D}  \left[\phi,
  \phi^\dagger, H \right] x^*_i x_i \e ^{-S_H} \,.
\end{equation}
Here $\phi=(x_i,\psi_i)^T, \phi^\dagger= (x_i^*,\bar\psi_i)$ is a
vector with $N$ bosonic and $N$ fermionic components. We set the
variance of $H$ to $g=1/2$. The action reads
\begin{equation}
  \label{eq:2}
  S_H = \I \phi^\dagger_i (H_{ij}-z\delta_{ij}) \phi_j + \tr
  H^2 \,.  
\end{equation}
Integration over $H$ yields an effective action
\begin{equation}
  S_4 = -\I z \phi^\dagger_i\phi_i +\frac14
  \trg(\phi_i\phi_i^\dagger)^2 \,.
\end{equation}
A Hubbard-Stratonovich transformation to $\Q-$variables (the $q,p$ are
bosonic and the $\theta, \bar\theta$ are fermionic),
\begin{equation}
  \Q = \begin{pmatrix} q & \bar\theta \\ \theta & \I p \end{pmatrix}
  \,,
\end{equation}
gives an action bilinear in $\phi$:
\begin{equation}
  S_\Q = \I \phi^\dagger_i (\Q-z) \phi_i + \trg \Q^2 \,.  
\end{equation}
Now we integrate over the $\phi-$fields {\it and} the
Grassmann-variables $\theta, \bar\theta$ exactly to get
an action, which depends only on the two real variables $q$ and
$p$. The DOS reads
\begin{widetext}
  \begin{equation}
    \langle n(z) \rangle = \frac{-2}{\pi N} \Im \int\!\! \mathcal{D}
     \left[\Q \right] \frac{q \e ^{-\trg\Q^2}}{\detg(\Q-z)^N} = \frac{-2}{\pi N}
    \Im \int\!\! \frac{\D p\, \D q}{\pi} q\e ^{-p^2-q^2} \frac{(\I
      p-z)^N}{(q-z)^N}\left( 1 - \frac{N}{2(q-z)(\I p-z)} \right) \,.
    \label{eq:n_int}
  \end{equation}
\end{widetext}
If we now set $z=E+\I\epsilon$ with $E\in\R$ and perform the
integration over $p$ and $q$, we get the exact finite$-N$ result \cite{mehta}
\begin{equation}
  \langle n(E)\rangle = \frac{\e ^{-E^2}}{2^N N! \sqrt\pi} \left(
    H_N^2(E) - H_{N+1}(E) H_{N-1}(E) \right) \,,
\end{equation}
where $H_n(E)$ denotes the $n-$th Hermite polynomial.
To perform an expansion around the limit $N=\infty$, we introduce
a rescaled energy variable $x=E/\sqrt{N}$. With this scaling the
band width becomes finite and the DOS a continuous normalizable
distribution in the limit $N\rightarrow \infty$. After the rescaling
(which is done for $p$ and $q$ as well) we get for the DOS
(after integration over $\theta,\bar\theta$ but before integration
over $p$ and $q$)
\begin{widetext}
  \begin{equation}
    \langle n(x) \rangle = \frac{-2}\pi \Im \int\!\! \frac{\D p\, \D
      q}{\pi/N} q\e ^{-Np^2-Nq^2} \frac{(\I p-x)^N}{(q-x)^N} \left( 1
      - \frac{1}{2(q-x)(\I p-x)} \right) \,,
    \label{eq:gueapp}
  \end{equation}  
\end{widetext}
which can be written as
\begin{equation}
  \langle n(x) \rangle = \int \D p\, \D q\, \e^{-NS(p,q,x)} f(p,q,x)
  \,.
\end{equation}
The large prefactor $N$ in the exponent allows for a
saddle-point approximation of the $p,q-$integrals in the usual way.
We get four solutions of the saddle-point equations
\begin{equation}
%  \begin{split}
    q_{\pm} = x/2 \pm\I/2\sqrt{2-x^2} \,, \qquad
    p_{\pm} = -\I q_\pm\,.
%  \end{split}
\end{equation}

\begin{figure}[htbp!]
    \includegraphics[scale=0.8]{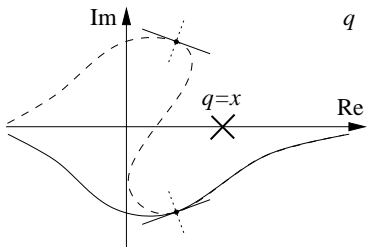} 
    \caption{\small The saddle-points of $q$ and possible
      integration paths. Solid lines denote the stable directions and
      dotted lines the unstable directions. 
%      It is impossible to go
%      through both
%      saddle-points without crossing a region where the integrand
%      is large.
             }
    \label{fig:spq}
\end{figure}

\begin{figure}[htbp!]
    \includegraphics[scale=0.8]{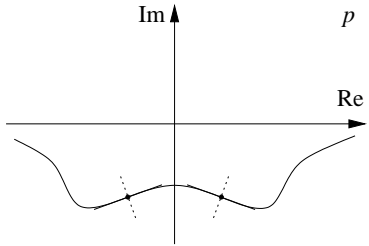}
    \caption{\small The saddle-points for $p$. 
%      It is possible
%      to choose a path of integration which reaches both
%      saddle-points.
             }
\label{fig:spp}
\end{figure}

\begin{figure}[htbp!]
    \includegraphics[width=7cm]{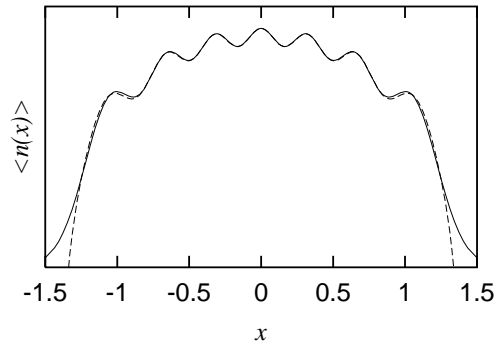} 
    \caption{\small Exact DOS of the GUE for $N=7$. 
      The dashed line is the $1/N-$approximation.}
    \label{fig:gue}
\end{figure}

The stability matrix is
\begin{eqnarray}
  \frac{\partial^2S}{2\partial q^2} &=&
  1-q_\pm^2 = \pm\I\sqrt{2-x^2} q_\pm =: \Delta_\pm
  \\
  \frac{\partial^2S}{2\partial p^2} &=&
  \Delta_\pm
  \\
  \frac{\partial^2S}{\partial p \partial q} &=& 0 \,.
\end{eqnarray}
Figures \ref{fig:spq} and \ref{fig:spp} show the location of the
saddle-points in the complex plane along with their stable directions
and possible integration paths.
The integral (\ref{eq:n_int}) converges only for
$\Im z>0$. This means, that the integration-path of $q$ must pass
below the singularity at $q=x$.  This forces us to use only the saddle-point
$q_-$ of the $q$-integration: A path of integration which also crosses
$q_+$ leads through the region between the two saddle-points, where
the integrand of (\ref{eq:gueapp}) is large. 
Therefore the correct choice for the path of integration over $q$
is the solid line in figure \ref{fig:spq}. In contrast it is possible
to use an integration-path which goes through both saddle-points for
the
$p-$integral, which is shown in figure \ref{fig:spp}.\\
After the change of variables
\begin{align}
  q &\rightarrow q_- +\frac{\delta q}{\sqrt{N\Delta_-}} & 
  p &\rightarrow p_\pm + \frac{\delta p}{\sqrt{N\Delta_\pm}}
\end{align}
we may perform an expansion in $\delta q$ and $\delta p$, which
defines
the $1/N-$expansion. The action $S^{(1)}_{SP}=S(q_-,p_-)$ vanishes and
yields the well-known semi-circle law as the only contribution
surviving in the $N\rightarrow\infty-$limit. The action
$S^{(2)}_{SP}=S(q_-,p_+)$ does not vanish but is purely imaginary and
yields
a contribution $\sim e^{-NS^{(2)}_{SP}}$ of modulus 1. This contribution
disappears in the $N=\infty-$limit, 
because the factor $(1-\frac{1}{2(q-x)(\I p-x)})$
vanishes for $q=q_-,p=p_+$. It yields, however, an important
oscillatory contribution $\propto 1/N$, whereas the 
(non-oscillatory) corrections to the first saddle-point $(q_-,p_-)$
start with order $1/N^2$.  
The final result is
\begin{equation}
  \langle n(x) \rangle = \frac{\sqrt{2-x^2}}{\pi} -
  \frac{(-1)^N \cos\left[NS_0(x)\right]}{\pi\sqrt 2 N (2-x^2)} + \mathcal{O}
  \left( \frac{1}{N^2} \right) \,
\end{equation}
with
\begin{align}
  S_0(x) &= S^{(2)}_{SP}(x) = x\sqrt{2-x^2} +2\arcsin(x/\sqrt2)
  \label{eq:k_phi1}
  \\
  \intertext{and}
  \label{eq:k_phi}
  S_0'(x) &= 2\sqrt{2-x^2} = 2\pi n_0(x) \,.
\end{align}
where the zeroth order approximation to the DOS (the semi-circle) is
written
as
$n_0(x)=(1/\pi)\sqrt{2-x^2}$.
The exact density of states and the $1/N-$ approximation  
 is shown in figure \ref{fig:gue}.
From (\ref{eq:k_phi1}) and
(\ref{eq:k_phi}) we note that the number of maxima of the
oscillatory $1/N-$term equals $N$ and the local width between
two maxima scales as the inverse of the DOS in zeroth approximation.
The positions of the maxima give therefore the
locations of eigenvalues for a ``typical'' realization of a random
matrix from the GUE with
one state per maximum on average.
Identifying the distance between to adjacent maxima with the
average level spacing at $x=E/\sqrt{N}$, we get for large
$N$ the following expression for $\Delta(E)$:
\begin{equation}
\Delta(E)\approx \frac{1}{\sqrt{N}n_0(E/\sqrt{N})} \,
\end{equation}
which is valid for $E$ not too close to the band edge. The
$1/N$-expansion diverges obviously at the band edge, because the
saddle-points coalesce and as a consequence the stability matrix
vanishes.

\section{ The other Ensembles}
\label{sec:three}
The elements of the orthogonal, 
symplectic and Class C/D ensembles are conveniently 
represented as $2N\!\times\!2N$-matrices having a $2\!\times\!2$-block
structure:
\begin{equation}
H=\begin{pmatrix}A&B\\C&D\end{pmatrix} \,,
\end{equation}
where the $N\times N-$matrices $A,B,C,D$ fulfill a set of conditions
defining the different ensembles. We set the variance of $H$ to $g=1$
in the following, which yields a bandwidth of 4 in all cases.

\subsection*{The orthogonal Ensemble}
The elements of the GOE are real symmetric $2N\!\times\!2N$-matrices:
$H=H^T=H^*$. To simplify calculations, it is convenient to use instead of $H$
the unitary equivalent matrix 
$M=UHU^\dagger$ with $U=\exp(\I(\pi/4)\sigma_1)\otimes\id_N$, where
$\sigma_1$ is a Pauli matrix acting on the $2\times2-$blocks.
If $H$ is real symmetric, $M$ fulfills the relation $M=\sigma_1 M^T\sigma_1$.
This entails the following relation among the block matrices
\begin{equation}
M=\begin{pmatrix}A&B\\B^\dagger&A^T\end{pmatrix},
\end{equation}
where $A$ is a hermitian $N\times N-$matrix and $B$ is complex symmetric.
As in  section II, we write for the DOS
\begin{equation}
  \langle n(z) \rangle = \frac{1}{2\pi N} \Re \int\!\! \mathcal{D}
   \left[\phi, \phi^\dagger, M \right] \; (x_{1i}^* x_{1i})
  \e ^{-S}  \,,
\end{equation}
with $\phi=(x_1,x_2,\psi_1,\psi_2)^T$, where the $x_{1/2},\psi_{1/2}$
are $N-$dimensional complex bosonic, resp. fermionic vectors and
$S$ reads
\begin{equation}
S=\I\phi^\dagger(M-z)\phi + \frac{1}{2}\tr M^2 \,.
\end{equation}
Now we introduce the notation
\begin{equation}
\chi_{1j}=(x_{1j},x_{2j}^*,\psi_{1j},\bar\psi_{2j})^T\,,\quad 
\chi_{2j}=(x_{2j},x_{1j}^*,\psi_{2j},\bar\psi_{1j})^T\,
\end{equation}
($j$ is the index of the $N-$dimensional tensor component) 
together with the transposition
\begin{equation}
\chi_{1j}^t=(x_{1j}^*,x_{2j},\psi_{1j}^*,-\psi_{2j})\,,\;
\chi_{2j}^t=(x_{2j}^*,x_{1j},\psi_{2j}^*,-\psi_{1j})\,.
\end{equation}
After Integration over $M$, we get the quartic action
in $\chi,\chi^t$:
\begin{equation}
  S_4 = -\I\chi_{1i}^t \chi_{1i} + \frac18 \trg
  (\chi_{1i}\chi_{1i}^t + \chi_{2i}\chi_{2i}^t)^2\,. 
\end{equation}
After transformation to the $\Q$-matrix
\begin{equation}
 \Q = 
  \begin{pmatrix}
    q_1          &  q_2^*     & \bar\theta_1 & -\theta_2 \\
    q_2          & q_1          & \bar\theta_2 & -\theta_1 \\
    \theta_1     & \theta_2     & \I p         & 0 \\
    \bar\theta_2 & \bar\theta_1 & 0            & \I p
  \end{pmatrix} 
\end{equation}
we obtain
\begin{equation}
  \langle n(z)\rangle= \frac{-1}{\pi N}\Im\int\!\! \mathcal D  \left[\Q \right]
  \frac{q_1 \e^{-\frac{1}{2}\trg\Q^2}} {\detg(\Q-z)^N}\,,
\end{equation}
with
\begin{multline}
  \detg\Q = \\
\frac{Q}{(\I p)^2} \exp \left( -\frac{2}{Q(\I p)} (
    q_1\Theta_1 - q_2\Theta_2 - q_2^*\bar\Theta_2 ) -
    \frac{2\Theta_2\bar\Theta_2}{Q(\I p)^2} \right)
\end{multline}
and
 $Q=q_1^2-q_2 q_2^*$, $\Theta_1=\bar\theta_1\theta_1 +
\bar\theta_2\theta_2$, $\Theta_2 = \bar\theta_1\theta_2$. Note
that the Grassmann-integration contains now quartic terms in
$\theta,\bar\theta$, which can nevertheless be done exactly.
The result is ($q=q_1,r=q_2 q_2^*$):
\begin{widetext}
  \begin{equation}
    \langle n(z)\rangle = \frac{-1}{\pi N} \Im \int\!\!  \frac{\D p \D
      q}{\pi} \D r\; \e ^{-p^2-q^2-r} q \frac{(\I
      p-E)^{2N}}{\left[(q-z)^2-r\right]^N} \left( 1 -
      \frac{2N(q-z)}{\left[(q-z)^2-r\right](\I p-z)} +
      \frac{(2N-1)N}{2\left[(q-z)^2-r\right] (\I p-z)^2} \right).
  \end{equation}
  Setting now as above $z=E+\I\epsilon$ and integrating over $p,q$ and
  $r$, we obtain the exact DOS in terms of Hermite-polynomials and the
  error-function:
  \begin{equation}
    \langle n(E)\rangle = \frac{\e ^{-E^2}}{4^NN\sqrt{\pi}} \left(
      \frac{H_{2N-1}(E)^2 - H_{2N}(E)H_{2N-2}(E)}{(2N-2)!}  -
      \frac{H_{2N-1}(E)}{(N-1)!}  \left( \sum_{k=1}^{N-1}
        \frac{2k!}{(2k)!} H_{2k-1}(E) - \sqrt{\frac\pi2} \e
        ^{\frac{E^2}2} \erf\frac E{\sqrt{2}} \right) \right)
  \end{equation}
\end{widetext}

\subsection*{The symplectic Ensemble}
Here we have
\begin{equation}
H=\begin{pmatrix}A&B\\B^\dagger&A^T\end{pmatrix}
\end{equation}
with $A$ hermitian and $B$ complex antisymmetric.
Proceeding as in the case of the GOE (the definition
of $\chi,\chi^t$ is the same), we find the quartic action
\begin{equation}
S_4 = \frac18 \trg (\chi_{1i}\chi_{1i}^t 
+ \Sigma_3\chi_{2i}\chi_{2i}^t\Sigma_3)^2, 
\end{equation}
where $\Sigma_3=\sigma_3\otimes\id_N$.
The corresponding $\Q-$matrix reads now
\begin{equation}
  \Q = 
  \begin{pmatrix}
    q        & 0            & \bar\theta_1 & \bar\theta_2 \\
    0        & q            & -\theta_2    & -\theta_1 \\
    \theta_1 & \bar\theta_2 & \I p_1       & \I p_2^* \\
    \theta_2 & \bar\theta_1 & \I p_2       & \I p_1
  \end{pmatrix} \,,
\end{equation}
and the DOS reads,
\begin{equation}
  \langle n(z)\rangle= \frac{-1}{\pi N}\Im\int\!\! \mathcal D  \left[\Q \right]
  \frac{q \e^{-\frac{1}{2}\trg\Q^2}} {\detg(\Q-z)^N}
\label{eq:symdos}
\end{equation}
with
\begin{multline}
  \detg\Q^{-1} = \\
  \frac{P}{q^2} \exp \left( \frac{2}{Pq} (
    \I p_1\Theta_1 - \I p_2\Theta_2 -\I p_2^*\bar\Theta_2 ) -
    \frac{2\Theta_2\bar\Theta_2}{Pq^2} \right)
\end{multline}
and $P=(\I p_1)^2 -(\I p_2)(\I p_2^*)$, $\Theta_1=\bar\theta_1\theta_1 +
\bar\theta_2\theta_2$ and $\Theta_2 = \bar\theta_2\theta_1$.
Setting $p=p_1$ and $r=p_2 p_2^*$, eq.(\ref{eq:symdos}) reads
after integration over the Grassmann variables, 
\begin{widetext}
  \begin{equation}
    \langle n(z) \rangle = \frac{-1}{\pi N} \Im \int\!\!
    \frac{\D p\D q}{\pi} \D r\; q \e ^{-p^2-q^2-r}
    \frac{\left[(\I p-z)^2+r\right]^N}{(q-z)^{2N}}
     \left( 1 - \frac{2N (\I
        p-z)}{\left[(\I p-z)^2+r\right] (q-z)} +
    \frac{(2N+1)N}{2\left[(\I p-z)^2+r\right] (q-z)^2} \right)
\end{equation}
and
\begin{equation}
    \langle n(E)\rangle =
    \frac{\e ^{-E^2}}{(2N)!\sqrt\pi} \left( \text{\rule{0em}{4ex}}
      \frac1{4^N} \left( H_{2N}(E)^2 - H_{2N+1}(E)H_{2N-1}(E) \right)
    - \frac{N!}{2N} H_{2N}(E)
      \sum_{k=0}^{N-1} \frac{H_{2k}(E)}{4^kk!}  \right) \,.
  \end{equation}
\end{widetext}

\subsection*{Class D}
The elements of class D are parameterized as
\begin{equation}
  H=\begin{pmatrix}A&B\\B^\dagger&-A^T\end{pmatrix} \,,
\end{equation}
with $A$ hermitian and $B$ complex antisymmetric.
The quartic action reads (in the notation above)
\begin{equation}
  S_4 = \frac18 \trg \left( (\chi_{1i}\chi_{1i}^t +
    \chi_{2i}\chi_{2i}^t)\Sigma_3 \right)^2 \,,
\end{equation}
and the $\Q$-matrix
\begin{equation}
  \Q = 
  \begin{pmatrix}
    q_1          &  q_2^* & -\theta_1    & -\theta_2 \\
    q_2          & -q_1     & \bar\theta_2 & \bar\theta_1 \\
    \bar\theta_1 & \theta_2 & \I p         & 0 \\
    \bar\theta_2 & \theta_1 & 0            & -\I p
  \end{pmatrix} \,.
\end{equation}

For $\langle n(z) \rangle$ we get
\begin{equation}
  \label{eq:41}
  \langle n(z)\rangle= \frac{-1}{\pi N}\Im\int\!\! \mathcal D  \left[\Q \right]
  \frac{q_1 \e^{-\frac{1}{2}\trg\Q^2}} {\detg(\Sigma_3\Q-z)^N} \,,
\end{equation}
with
\begin{multline}
  \detg\Sigma_3\Q = \\
  \frac{Q}{(\I p)^2} \exp \left( -\frac{2}{Q(\I p)} (
    q_1\Theta_1 - q_2\Theta_2 + q_2^*\bar\Theta_2 ) -
    \frac{2\Theta_2\bar\Theta_2}{Q(\I p)^2} \right)
\end{multline}
and $Q=q_1^2+q_2 q_2^*$, $\Theta_1=\bar\theta_1\theta_1 -
\bar\theta_2\theta_2$, $\Theta_2 = \theta_1\theta_2$.  The integrand
of (\ref{eq:41}) contains a pole depending on $q=q_1$ and
$r=q_2q_2^*$. To circumvent it, we have to deform the integration
path into the complex plane:
\begin{align}
  q &\rightarrow \e ^{-\I \pi/4} q + \epsilon
  \e ^{\I \pi/4} \tanh(q/\epsilon)
  \\
  \sqrt{r} &\rightarrow \e ^{\I \pi/4} \sqrt{r} + \epsilon
  \e ^{-\I \pi/4} \tanh(\sqrt{r}/\epsilon)
\end{align}
for $\Re z>0$ and $\epsilon$ must be small enough: $0<\epsilon<(\Im z
+ \Re z)/\sqrt2$.  We find then
\begin{widetext}
  \begin{equation}
%  \begin{split}
    \langle n(z) \rangle = \frac{-1}{\pi N} \Im \int\!\!
    \frac{\D p\,\D q}{\pi} \D r\;
    q \e ^{-p^2-q^2-r}
    \frac{(\I p-z)^{2N}}{\left[(q-z)^2+r\right]^N}
%    \\
%    &\quad{} \hspace{11em} \times 
      \left( 1-
      \frac{N(2N-1)}{2\left[(q-z)^2+r\right](\I p-z)^2} \right)
%  \end{split}
\end{equation}
and
\begin{equation}
% \begin{split} 
    \langle n(E)\rangle =
    \frac{-(-1)^N \e ^{-E^2}}{N! 4^N \sqrt\pi}
    \left( \rule{0em}{4ex}
      \frac{(-1)^N(N-1)!}{(2N-2)!} H_{2N}(E) H_{2N-2}(E)
%    \right.
%    \\
%    &\quad\left. \hspace{3.5em} 
     + 2 E H_{2N-1}(E) \left( 
        \sum_{k=1}^{N-1}\frac{(-1)^kk!}{(2k)!}H_{2k}(E)+1+\frac1{2E^2}
      \right) 
    \right) \,.
%  \end{split}
\end{equation}
\end{widetext}

\subsection*{Class C}
The parameterization reads
\begin{equation}
  H=\begin{pmatrix}A&B\\B^\dagger&-A^T\end{pmatrix}
\end{equation}
with $A$ hermitian and $B$ complex symmetric.
The quartic action is:
\begin{equation}
  S_4 = \trg (\chi_{1i}\chi_{1i}^t\Sigma_3 +
  \Sigma_3\chi_{2i}\chi_{2i}^t)^2 \,,
\end{equation}
with the $\Q-$matrix
\begin{equation}
  \begin{pmatrix}
    q            & 0            & -\theta_1 & \bar\theta_2 \\
    0            & -q           & -\theta_2 & \bar\theta_1 \\
    \bar\theta_1 & \bar\theta_2 & \I p_1    & \I p_2^* \\
    \theta_2     & \theta_1     & \I p_2    & -\I p_1
  \end{pmatrix} \,. 
\end{equation} 
We have
\begin{equation}
  \langle n(z)\rangle= \frac{-1}{\pi N}\Im\int\!\! \mathcal D \left[\Q
  \right] \frac{q_1 \e^{-\frac{1}{2}\trg\Q^2}} {\detg(\Sigma_3\Q-z)^N}
  \,,
\end{equation}
with
\begin{multline}
  \detg(\Sigma_3\Q)^{-1} =  \\
  \frac{P}{q^2} \exp \left( \frac{2}{Pq} (
    \I p_1\Theta_1 + \I p_2\Theta_2 -\I p_2^*\bar\Theta_2 ) -
    \frac{2\Theta_2\bar\Theta_2}{Pq^2} \right)
\end{multline}
and  $P=(\I p_1)^2 +(\I p_2)(\I p_2^*)$, $\Theta_1=\bar\theta_1\theta_1 -
\bar\theta_2\theta_2$ and $\Theta_2 = \bar\theta_1\bar\theta_2$.
With $p=p_1,r=p_2 p_2^*$ the result reads
\begin{widetext}
  \begin{align}
    \langle n(z)\rangle &= \frac{-1}{\pi N} \Im \int\!\!  \frac{\D p\,
      \D q}{\pi} \D r\; \e ^{p^2-q^2-r} q \frac{\left[(\I p-z)^2-r
      \right]^N} {(q-z)^{2N}}
    \left( 1-
      \frac{N(2N+1)}{2\left[(\I p-z)^2-r\right](q-z)^2} \right) \,,
    \\
    \langle n(E) \rangle &= \frac{(-1)^NN!\e ^{-E^2}}{(2N)!\sqrt\pi}
    \left(
      H_{2N}(E) \sum_{k=0}^N \frac{(-1)^kH_{2k}(E)}{k!4^k}  
      + \frac{H_{2N+2}(E)}{4N} \sum_{k=0}^{N-1} \frac{(-1)^k
        H_{2k}(E)}{k!4^k} 
    \right) \,.
  \end{align}
\end{widetext}

\section{ Saddle-Point Approximation}
\label{sec:four}
The saddle-point approximation for the four ensembles
proceeds as in the unitary case in general: All four
ensembles share the saddle-points
$q_\pm =x/2 \pm \I\sqrt{1-x^2/4}$, $p_\pm = -\I q_\pm$, $r_0=0$,
($x=E/\sqrt{N}$).
As in the unitary case, the points $(q_+,p_\pm,r_0)$ have
to be avoided by the integration path. The fluctuations are
quadratic in $p$ and $q$ and linear in $r$ around the
two remaining SP's $(q_-,p_\pm,r_0)$. Apart from them, we have
additional
SP's for the GOE and the GSE, given in Table \ref{tab:tab}.\par
\begin{table}[htbp]
\caption{The additional saddle-points for the GOE and the GSE}
\label{tab:tab}
\begin{ruledtabular}
\begin{tabular}{lCCC}
  Ensemble & p & q & r
  \\
  GOE & -\I\left(\frac x2 \pm\I\sqrt{1-\frac{x^2}4}\right) & \frac x2
  & \frac{x^2}4 -1 
  \\ 
  GSE & -\I\frac x2 & \frac x2 \pm\I\sqrt{1-\frac{x^2}4} &
  1-\frac{x^2}4  
\end{tabular}
\end{ruledtabular}
\end{table} 
The additional SP can not be reached in the GOE case $(r_{SP}<0)$ but
has to be included for the GSE $(r_{SP}>0)$.  In class C and D there
is an additional saddle-point manifold at $x=0$. However, our results
displayed below are obtained by confining the analysis to the standard
saddle-points, which exist for all values of $x$, thereby assuming
continuity in the limit $x\rightarrow 0$.  In principle this procedure
could have led to an additional singularity at $x=0$ similar to the
divergence at the band edge.  But the comparison with the exact DOS in
these cases shows that our assumption is indeed correct and the
additional saddle-points play no role in computing the
$1/N-$expansion.

We give in the following the results for the $1/N$-expansion of
the GOE, GSE and  Class C/D.

\subsection*{The orthogonal Ensemble}

Lets define the $N=\infty$ approximation to the DOS of the GOE
as $n_0(x)=(1/\pi)\sqrt{1-x^2/4}$. Then the $1/N-$expansion reads to
order $1/N^2$
\begin{widetext}
  \begin{equation}
    \langle n(x)\rangle = n_0(x)
    -\frac{1}{N}\frac{1}{8\pi^2n_0(x)}+\frac{1}{N^2}
    \frac{3+x^2}{128\pi^6n_0(x)^5}
    -\frac{1}{N^2}\frac{1}{64\pi^6n_0(x)^5}
    \cos\left[2NS_0(x)-\arcsin(x/2)\right] \,,
\end{equation}
\end{widetext}
with
\begin{eqnarray}
  S_0(x) &=& 2\arcsin(x/2)+x\sqrt{1-x^2/4},\\
  S_0'(x) &=& 2\pi n_0(x) \,,
\end{eqnarray}
analogous to the unitary case. We observe the same features as in the
GUE case: The number of maxima of the oscillatory part within the band
$-2<x<2$ given by the semi-circle is $2N$, the total number of levels,
so we have one level per maximum. The correction diverges close to the
band edge as expected. Apart from the oscillatory contribution there
is a non-oscillatory contribution of order $1/N$. However, the
information about the level repulsion is encoded in the oscillating
part $\propto \cos(2NS_0(x))$.  This term is of order $1/N^2$, which
means that the repulsion of levels is weaker in the GOE than in the
GUE.  Formally we can write
\begin{eqnarray}
  \langle n(x)\rangle &=& n_0(x) +
  \frac{1}{N}\left({\textrm{non-oscillating term}}\right) \nonumber 
  \\
  &&{} +\frac{1}{N^\alpha}\left({\textrm{oscillating term}}\right)
\end{eqnarray}
with $\alpha=2$. For the GUE $\alpha=1$. Now the two-level correlation
function $R_2(r)$ behaves for the Wigner-Dyson Ensembles as\cite{mehta}:
\begin{equation}
R_2(r)\sim r^\beta\ \ \ \ {\textrm{for}}\ \ \ \ r\ll 1 \,,
\end{equation}
where $r$ is the distance between two levels on a scale corresponding to 
the average level spacing $\Delta(x)\sim 1$. For the GUE we have $\beta=2$ and for
the GOE $\beta=1$.
We conjecture therefore
\begin{equation}
\beta=\frac{2}{\alpha}
\label{eq:conj}
\end{equation}
as a relation between the universal parameter $\beta$, which characterizes
the short distance behavior of the two-level correlation function
and the exponent $\alpha$ of the factor $1/N$, which multiplies
the oscillating term in the $1/N-$expansion. With this relation we can
extract from the $1/N-$expansion of the one-point function universal
information about the two-point function, which discerns the
WD-Ensembles from each other (whereas the DOS is the same for
all three ensembles in the limit $N=\infty$).

\subsection*{The symplectic Ensemble}
We have (with the same definition of $n_0(x)$ and $S_0(x)$ as in the GOE)
\begin{eqnarray}
  \langle n(x)\rangle &=& n_0(x) + \frac{1}{N}\frac{1}{8\pi^2n_0(x)}
  \nonumber
  \\
  &&{} - \frac1{N^{\frac{1}{2}}} \frac{
  \cos\left[NS_0(x)+\frac12\arcsin(x/2)\right] } {
    (-1)^N \sqrt8\pi^{\frac{3}{4}}n_0(x)^{\frac{1}{4}}} \,.
\end{eqnarray}
Because each eigenvalue appears twice in a $2N${}$\times${}$2N$-matrix from
the symplectic ensemble due to Kramers degeneracy \cite{mehta}, we
have only $N$ different eigenvalues and therefore $N$ maxima within
the band. 
%%%%%%%%%%%%%%%%%%%%%%%%%%%%%%
\begin{figure*}[htbp!]
    \includegraphics[width=7cm]{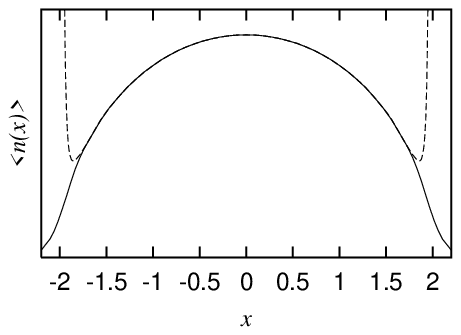} 
%    \caption{\small The GOE.}
%    \label{fig:goe}
%\end{figure}
\hspace{2cm}
%\begin{figure}[htbp!]
    \includegraphics[width=7cm]{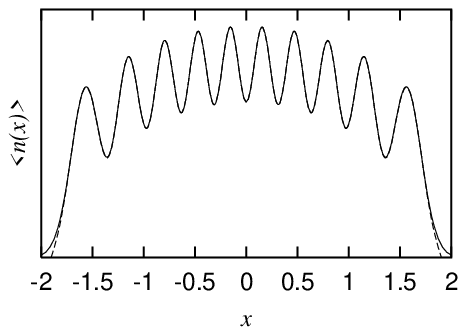} 
    \caption{\small The GOE and the GSE. $N=10$ in both cases. The
      solid line is the exact DOS, the dashed line is the saddle-point
      approximation.}
    \label{fig:goegse}
\end{figure*}
%%%%%%%%%%%%%%%%%%%%%%%%%%%%%%%%%%%%%%%
The oscillatory term comes with a $1/\sqrt{N}$ prefactor,
therefore $\alpha=1/2$, which gives with (\ref{eq:conj}) the correct
universal exponent $\beta=4$.

%%%%%%%%%%%%%%%%%%%%%%%%%%%%%%%%
\begin{figure*}[htbp!]
    \includegraphics[width=7cm]{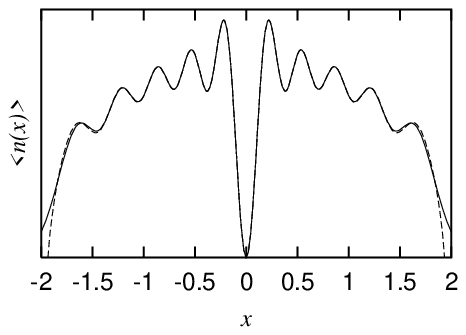}
%    \caption{\small Class C.}
%    \label{fig:clc}
%\end{figure}
\hspace{2cm}
%\begin{figure}[htbp!]
    \includegraphics[width=7cm]{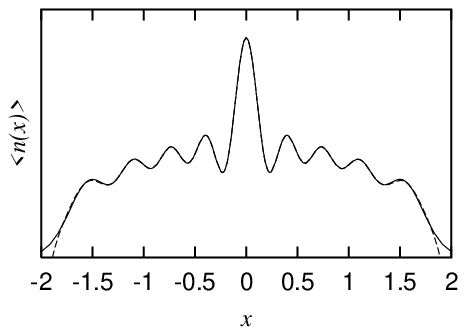} 
    \caption{\small Class C and Class D. $N=5$ in both cases.The
      solid line is the exact DOS, the dashed line is the saddle-point
      approximation.}
    \label{fig:clcd}
\end{figure*}
%%%%%%%%%%%%%%%%%%%%%%%%%%%%%%  

\subsection*{Class C}
The $1/N-$expansion yields
\begin{eqnarray}
  \langle n(x)\rangle &=&
  n_0(x) + \frac{1}{N}\frac{1}{8\pi^2n_0(x)} \nonumber \\ 
  &&{} - \frac{1}{N}
\frac{\sin\left[2NS_0(x)+\arcsin(x/2)\right]}{4\pi^3xn_0(x)^2}
\label{lNclC} \,.
\end{eqnarray}
Each element of the Class C ensemble has a spectrum symmetric with
respect to $x=0$: To each eigenvalue $x$ there is a state with
eigenvalue $-x$. No matrix from Class C has an eigenvalue zero.
Therefore the DOS vanishes at $x=0$. The number of maxima in the band
equals $2N$, we have again one (non-degenerate) level per maximum. Now
the spectral region around $x=0$ is different from the WD-Ensembles
because there is no smooth limit of $\langle n(x)\rangle$ for
$N\rightarrow\infty$: If one introduces the rescaled variable
$y=2Nx/\pi$ and considers the microscopic limit $x\rightarrow 0$,
$N\rightarrow\infty$, keeping $y$ finite, the DOS reads
\begin{equation}
\langle n(y)\rangle = \frac{1}{\pi}\left(1-\frac{\sin(2\pi
    y)}{2\pi y}
\right) \,,
\end{equation}
which coincides with the result in [\onlinecite{altz}]. In terms of
$y$, the correction term is independent of $N$ in contrast to the GUE,
where the oscillatory contribution to the DOS vanishes as $1/N$ for
large $N$. This, however, is not a signal of an enhanced level
repulsion close to $x=0$ in Class C. It is merely due to the lack of
(approximate) translational invariance in the band center, which is
caused by the mirror-symmetry mentioned above. In the GUE,
translational invariance is broken only through the term proportional
to $\exp(-\sum^N_{i=1}\lambda_i^2)$ in the joint probability
distribution of the eigenvalues $\lambda_i$, which can be neglected
close to the band center and large $N$. Therefore the oscillatory
structure in $\langle n(x)\rangle$ is smeared out by the summation
over all members of the ensemble, whereas in Class C the quotient of
the probability to find a level at the first maximum respective the
first minimum of $\langle n\rangle$ is independent of $N$. The
spectrum is therefore more rigid near $x=0$. The two-level correlation
function $R_2(y_1,y_1+y)$ behaves for $0<y_1,y\ll 1$ nevertheless as
$y^2$ and belongs therefore to the unitary universality
class\cite{altz}. The correction term does not modify the macroscopic
behavior of $\langle n\rangle$ because the function $\sin(Nx)/(Nx)$
tends to zero in the $L_2-$sense for $N\rightarrow\infty$.  That means
that the probability to find $m$ states in a region of width $\Delta
x=\frac{\pi m}{2N}$ around zero tends to one for $1\ll m\ll N$ in the
limit $N\rightarrow\infty$, as in the WD-Ensembles.

Because the $1/N$-approximation is reliable everywhere except at the
band edge, we can compute the average level spacing exactly in
the vicinity of $x=0$ without recourse to measure $x$ (respective $y$)
in units of the mean spacing at a distance of many spacings
from zero\cite{altz}. In terms of $y$ (which is {\it exactly} related
to $x$ and the original variable $E$ through $N$), the average spacing 
between the first and second level $>0$ is given by
\begin{equation}
\langle\Delta_y(1,2)\rangle = \frac{1}{2\pi}(z_4-z_2) \approx 1 \,,
\end{equation}
where $z_k>0$ denotes the $k-$th zero of $\tan(z)-z$.
This relation is valid for large $N$ but $\Delta(j,j+1)$ can be
computed via (\ref{lNclC}) for all levels $j$ and $j+1$
not too close to the band edge and arbitrary $N$.

\subsection*{Class D}
The same mirror-symmetry as in Class C is valid in Class D.
We have
\begin{eqnarray}
  \langle n(x)\rangle &=&
  n_0(x) - \frac{1}{N}\frac{1}{8\pi^2n_0(x)} \nonumber \\
  &&{} + \frac{1}{N}
  \frac{\sin\left[2NS_0(x)-\arcsin(x/2)\right]}{4\pi^3xn_0(x)^2}
\end{eqnarray}  
very similar to Class C, but now the DOS is enhanced in the band
center because there are always two states with eigenvalue close to
zero.  The number of different maxima is $2N-1$. Again this feature
vanishes in the $N=\infty-$limit, because the single additional state
at $x=0$ has measure zero for vanishing average level distance
$\Delta(E)\sim 1/N$.  As in the Class C ensemble we have $\alpha=1$
and the universal exponent is $\beta=2$. Class D belongs therefore
with Class C to the unitary universality class, which determines the
level repulsion even in the immediate vicinity of $E=0$ e.g. for the
second and third state away from zero.

Figures \ref{fig:goegse} and  \ref{fig:clcd} give the exact DOS and the
$1/N$-approximation (dashed lines) for the four Ensembles.

\section{Conclusions}
\label{sec:concl}
We have computed the exact density of states for the three
Wigner-Dyson Ensembles as well as for Class C and Class D by
evaluating finite dimensional
supersymmetric integrals analytically. In this way the 
$N\! \!\rightarrow\!\!\infty$-limit implicit in most of the previous
calculations could be avoided. The exact results were then
employed to test a $1/N-$expansion, which differs from the
usual one because it proceeds only in the bosonic sector
whereas the fermionic sector is evaluated exactly. The
$1/N-$expansion therefore does {\it not} start from a supersymmetric
saddle-point or saddle-point manifold. 
It turned out that for all ensembles
only a discrete set of saddle-points is important, whereas the
saddle-point manifold at $x=0$ appearing in Class C/D is not
needed for an almost exact computation of the DOS everywhere
in the spectrum (including $x=0$) except the band edge. 
The $1/N-$expansion revealed a connection between the
order $\alpha$ of $N^{-1}$ multiplying the oscillatory term
of the DOS and the universal short distance exponent $\beta$
of the two-level correlation function, eq.(\ref{eq:conj}). 
In this way we obtain information about the two-point
function from the one-point function.
We believe that
relation (\ref{eq:conj}) is always fulfilled, if the spectrum is 
approximately ``translational invariant'', i.e. $R_2(x_1,x_2)$ depends only
on $x_2-x_1$ (this is implicit in all calculations using an
unfolding procedure). At points in the spectrum where this
invariance is broken (as at $x=0$ for Class C/D) the two-point
function has to be calculated itself, which can, of course, also be
done with our method. 

The $1/N-$expansion of the DOS 
yields an unambiguous
determination of the average level spacing $\Delta(E)$ in terms of $E$ and  
$N$ everywhere in the spectrum including $x=0$ for all ensembles.
For Class C/D we conclude that the special features of the DOS at
the band center vanish in the macroscopic $N\rightarrow\infty-$limit
and are of no relevance if the bandwidth is kept finite. 

The other Gaussian ensembles are under current investigation. It
should also be possible to extend the method to non-Gaussian
matrix models, see e.g. [\onlinecite{ake}], which are difficult to
treat with orthogonal polynomials. Here it would be interesting to
test the results obtained in the large $N-$limit using
singular integral equations by comparing with the 
exact formulae for finite $N$ 
and the systematic $1/N-$expansion.

\begin{acknowledgments}
  This work was supported by the \emph{Graduiertenkolleg} ``Nonlinear
  Problems in Analysis, Geometry, and Physics'' (GRK 283), financed by
  the German Science Foundation (DFG) and the State of Bavaria.
\end{acknowledgments}

\end{document}